\documentclass[aps,prl,showpacs,showkeys,superscriptaddress,twocolumn]{revtex4}

\usepackage{graphicx}%

\usepackage[intlimits]{amsmath}
\usepackage{amsfonts}
\usepackage{amsthm}
\usepackage{mathrsfs}
\usepackage{stmaryrd}

\begin{document}


\title{Importance of hydrodynamic shielding for the dynamic behavior of short 
polyelectrolyte chains}



\author{Grass, Kai}
\email{grass@fias.uni-frankfurt.de}
\affiliation{Frankfurt Institute for Advanced Studies, JW Goethe Universit\"at, Ruth-Moufang-Strasse 1,
D-60438 Frankfurt, Germany}
\author{B{\"o}hme, Ute}
\author{Scheler, Ulrich}
\affiliation{Leibniz Institute for Polymer Research, Hohe Str. 6,
D-01069 Dresden, Germany}
\author{Cottet, Herv\'{e}}
\affiliation{Institut des Biomol\'{e}cules Max Mousseron, (UMR 5247 CNRS -
Universit\'{e} de Montpellier 1 - Universit\'{e} de Montpellier 2), 2, place
Eug\`{e}ne Bataillon CC 017, 34095 Montpellier Cedex 5, France}
\author{Holm, Christian}
\email{c.holm@fias.uni-frankfurt.de}
\affiliation{Frankfurt Institute for Advanced Studies, JW Goethe Universit\"at, Ruth-Moufang-Strasse 1,
D-60438 Frankfurt, Germany}
\affiliation{Max Planck Institute for Polymer Research, Ackermannweg 10, D-55128
Mainz, Germany}

\date{\today}

\begin{abstract}
  The dynamic behavior of polyelectrolyte chains in the oligomer
  range is investigated with coarse-grained molecular dynamics
  simulation and compared to data obtained by two different
  experimental methods, namely capillary electrophoresis and
  electrophoresis NMR.  We find excellent agreement of experiments and
  simulations when hydrodynamic interactions are accounted for in the
  simulations.  We show that the electrophoretic mobility exhibits a
  maximum in the oligomer range and for the first time illustrate that
  this maximum is due to the hydrodynamical shielding between the
  chain monomers.  Our findings demonstrate convincingly that it is
  possible to model dynamic behavior of polyelectrolytes using coarse
  grained models for both, the polyelectrolyte chains and the solvent
  induced hydrodynamic interactions.
\end{abstract}


\maketitle

\section{Introduction}

Electrophoresis methods are widely used to separate
(macro)biomolecules \cite{righetti96a,dolnik06a} such as peptides,
proteins, and DNA, as well as synthetic polymers
\cite{cottet05a,cottet07a}.  Short polyelectrolytes (PEs) can conveniently
be separated in free solution without the aid of a gel by capillary
electrophoresis (CE).  Additionally, CE is employed to characterize
the hydrodynamic properties of charged biomolecules, in particular the
electrophoretic mobility, $\mu$, the diffusion coefficient, and the hydrodynamic
radius \cite{nkodo01a,stellwagen02a}.

Alternatively, these properties can be determined by pulsed field
gradient (PFG) NMR \cite{stejskal65a,stilbs06a}. With a combination of
diffusion NMR and electrophoretic NMR, the charge of macromolecules
\cite{scheler02a}, the influence of ionic strength
and the dielectric constant of the solution have been
investigated \cite{boehme03-07}.

Several studies of PEs of well defined length in the short chain
regime have shown that the free-solution mobility $\mu$ exhibits a
characteristic behavior
\cite{hoagland99a,cottet00b,stellwagen02a,stellwagen03a}: from the
monomer mobility, $\mu_0$, onwards, $\mu$ increases towards a maximum,
$\mu_{\text{max}}$, that occurs for chains of a specific degree of
polymerization, $N_{\text{max}}$. After this maximum, $\mu$ decreases
slightly to reach a constant value, $\mu_\text{FD}$, the so-called
free draining mobility. Very little is known about the origin of the
maximum, yet the knowledge of the precise dynamical behavior of a PE
is a prerequisite for designing possible applications in microfluidic
devices, such as electrophoretic separation or gene sequencing.

Whether or not a mobility maximum appears and the precise value of
$N_{\text{max}}$ seems to depend on the flexibility of the
chain. While a maximum has been observed for flexible PEs
such as single stranded DNA or sulfonated polystyrene (PSS) at values
around 10 repeat units for
$N_{\text{max}}$\cite{hoagland99a,cottet00b,stellwagen03a},
semi-flexible double-stranded DNA does not exhibit it. The small
number for $N_{\text{max}}$ and the strong influence of the
persistence length shows that a short chain behavior must be
responsible for this. Furthermore, it has been shown that the
$\mu_{\text{max}}$ is diminished under elevated salt conditions.

Existing theories 
\cite{muthukumar96a,volkel95b,mohanty99a,hoagland99a,stellwagen03a}
have been successful in describing the qualitative behavior of an
initially rising mobility as well as the constant long-chain limit,
but they have not been able to reproduce this maximum or explain its
origin.

To provide a fundamental understanding of the involved dynamics of
PEs, we propose a coarse-grained Molecular Dynamics
model that takes all charged particles (i.e. PE repeat
units, counter ions, and additional salt) and electrostatic and
hydrodynamic interactions (HI) between them into account. In the following
we demonstrate that this model \emph{quantitatively} reproduces
experimental results obtained by two completely different experimental
techniques. In addition we suggest a microscopic interpretation of the
size-dependence of $\mu$ for short
PE chains, based on hydrodynamic shielding, which gives
fundamental insight into the interplay of hydrodynamic friction and
charge correlations for charged macromolecules.

\section{Simulations}

\textbf{Model.} We simulate a flexible PE using a
bead-spring model employing the ESPResSo package
(\cite{limbach06a}). All parameter values are chosen to match the
properties of PSS used in the experiments and given in reduced units with $k T = 1.0$ and
$\sigma_0 = 2.5 $A being the energy and the relevant length scale.
The beads (chain monomers) are connected by FENE bonds with stiffness
$k = 30$, and maximum extension $R = 1.5$. Additionally a truncated
Lennard-Jones or WCA potential with depth $\epsilon = 0.25$ and width
$\sigma = 1$, is used for excluded volume interactions. Each monomer
has a charge of $q=-1$ in units of $e$. Monovalent counter ions,
$q=1$, and monovalent positive and negative salt ions are subjected to
the same LJ potential giving all particles the same size. The
simulations are carried out under periodic boundary conditions in a
rectangular simulation box of size $L=34$ (for $N=2$) to $L=89$ for
($N=32$) resulting in constant monomer concentration of approximately 1 g/l or 5
mM. All electrostatic interactions are calculated with the P3M algorithm. 
The Bjerrum length $l_B = {e^2}/{4 \pi \epsilon k T} = 2.84$ in
simulation units corresponds to 7.1 A (the Bjerrum length for water at
room temperature). Together with the model's average bond length of
$b=0.91$, we compute a Manning factor of $\xi=l_B/b=3.12$. The
inclusion of HI is done via frictional coupling
of the beads to a Lattice Boltzmann (LB) fluid as detailed in
\cite{ahlrichs99a}. The modeled fluid has a kinematic viscosity $\nu =
3.0$, a fluid density $\rho = 0.864$, and is discretized by a grid
with spacing $a=1.0$. The coupling parameter is $\Gamma_{\text{bare}}
= 20.0$. The simulation time step is $\tau = 0.01$.

To determine the impact of HI, we compare the
results to simulations with a simple Langevin thermostat that does not
recover long-range hydrodynamic interactions between the monomers, but
only offers local interaction with the solvent.

\textbf{Transport coefficients.} We determine two different transport
coefficients for the model PE
that are likewise determined in the associated experiments.
The single chain diffusion coefficient is obtained from the center of mass
velocity auto correlation function:
$ D = \frac{1}{3} \int_{0}^{\infty} d \tau \langle \vec{v_c}(\tau) \vec{v_c}(0) \rangle$.

In CE experiments, the electrophoretic
mobility $\mu$ of the solute is determined by
$ \mu = {v}/{E} = {L l}/{V t} $,
where $v$ is the velocity, $E$ is the electric field, $V$
is the applied voltage, $L$ is the total length of the capillary, $l$ is the
migration (or effective) length up to the detector and $t$ is the detection time
of the solute. 

In the simulations, we use a Green-Kubo relation to obtain $\mu$ at
zero electric field. This approach has been successfully applied in
simulations to determine the electrophoretic mobility of charged
colloids (\cite{lobaskin07a}). The chain mobility is calculated from
the correlation function between the center of mass velocity of the
polyelectrolyte chain and the velocities of all charged particles in
the system: $\mu = \frac{1}{3 k_B T} \sum_i q_i
\int_{0}^{\infty}\langle \vec{v}_i(0) \cdot \vec{v_c}(\tau) \rangle
d\tau$. This method guarantees that no conformational changes of the
chain structure or the ion distribution are induced by an artificial
high external field, which is sometimes used in other simulations to
separate the directed electrophoretic motion from Brownian motion
within reasonable computing time. Our method enables us furthermore to
obtain both transport properties from the same simulation trajectories
without additional computational effort.\footnote{The correlation
  functions were integrated using an analytic fit for the long-time
  tail. The associated uncertainties are included in the error bars.}

\section{Experiments}

\textbf{Capillary electrophoresis (CE)} is an analytical separation
technique based on the differential migration of ionic species under
electric field \cite{righetti96a}.

The CE experiments were performed using an Agilent technologies
capillary electrophoresis system (Agilent, Waldbronn, Germany). 
Capillaries of 33.5 cm (25 cm to the detector) length,
and 50 $\mu$m diameter were prepared from bare silica tubing purchased from Supelco
(Bellefonte, PA, USA). New capillaries were conditioned by performing the
following flushes: 1M NaOH for 30 min, 0.1 M NaOH for 10 min, and water for 5
min. Samples were introduced hydrodynamically ($\sim 4$ nL) at 0.5 g/L 
concentration ($\sim
2.5$ mM monomer concentration). The electrolyte was pure water. Solutes were
detected at 225 nm. The electric field was
kept constant at 224 V/cm (V=+7.5 kV). The polarity of the applied voltage on
the inlet side of the capillaries was positive. All the experiments were
performed at 27~$^\circ$C. Electro osmotic mobilities were determined from the
migration time of a neutral marker (mesityl oxide, $\sim 0.1$\% (v/v) in the
electrolyte).            
The sodium polystyrene sulfonate standards ($M_w 1.430 \times 10^3, 5 \times
10^3, 8 \times 10^3, 16 \times 10^3, 
31 \times 10^3, 41 \times 10^3, 88 \times 10^3, 177 \times 10^3, 350 \times
10^3$; $M_w/M_n$ ca 1.1) were purchased from
American Polymer Standards Corp. (Mentor, OH, USA). These standards have
verified sulfonation rates larger than 88\% and are almost fully charged.
Purified water delivered by an Alpha-Q system (Millipore, Molsheim, France) was
used to prepare all electrolytes and sample solutions.

\textbf{In electrophoretic NMR,} diffusion and electrophoretic motion
are separated by the design of the experiment \cite{gottwald03a}.  No
gel has been used, so that self diffusion and free electrophoresis are
measured.  The PSS samples have been obtained from Fluka. To minimize
the effects of variations of the ionic strength \cite{boehme03-07},
samples have been dialyzed against water (cut off volume 0.5ÊkDA), and
subsequently dried under vacuum. For all experiments a monomer concentration
of 5 mM in deuterated water has been used. The diffusion
experiments have been performed on a Bruker Avance 500 NMR
spectrometer operating at a Larmor frequency of 500ÊMHz for protons
equipped with a DIFFÊ30 probe head generating a maximum pulsed field
gradient strength of 12ÊT/m. The gradient pulse duration and
diffusion times have been adjusted between 0.8 to 2Êms and 8 to 20Êms respectively for optimal
resolution for each molecular weight resulting in different diffusion
coefficients. Because of their narrow molecular weight distribution,
diffusion coefficients have been determined by a linear fit to the
Stejskal-Tanner equation \cite{stejskal65a}.  Electrophoresis NMR
experiments have been performed on a Bruker Avance 300 NMR
spectrometer operating at a Larmor frequency of 300ÊMHz for protons
with an in-house-built electrophoresis probe head utilizing a Bruker
micro2.5 imaging gradient system generating magnetic field gradient
strength of up to 1ÊT/m. The flow times varied between 15 and
50 ms and the gradient pulse duration between 3 and 6 ms
respectively with gradient amplitudes between 0.3 and 0.6 T/m. The
electric field has been linearly incremented between -140 V/cm and
+140 V/cm. The electrophoretic mobility is a model-free read out from
the two-dimensional electrophoresis NMR spectrum correlating the
chemical shift, that identifies the moving species, with
electrophoretic mobility \cite{scheler02a}.

\section{Discussion}

\begin{figure}[htp]
\begin{center}
  \includegraphics[width=\columnwidth]{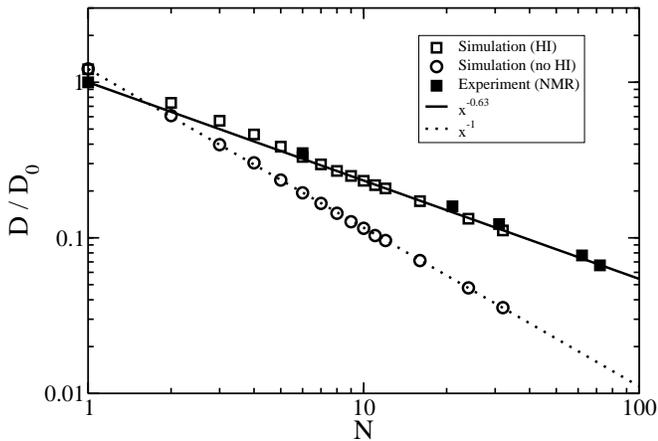}
  \caption[]{\label{fig:diffusion}The normalized diffusion coefficient, $D/D_0$,
  for PSS of different
  lengths $N$. The diffusion for simulations with HI and the experimental data
  agree and show a scaling exponent of $m=0.63$. Without HI the diffusion scales with
$m=1$.}
\end{center}
\end{figure}

\textbf{Diffusion.} Recently, Stellwagen et~al. \cite{stellwagen03a} observed
that the diffusion coefficient, $D$, of PSS can be approximated by a power law
scaling $D = D_0 N^{m}$, where $D_0$ is the monomer diffusion coefficient, and
$m$ is the scaling exponent. 
In Figure \ref{fig:diffusion}, we compare the diffusion coefficient
obtained from simulations to the results from the NMR study. 
The simulated data is normalized by $D_0 = 0.052$ as obtained from the power law
fit, and  the experimental data by the monomer diffusion coefficient of $D_0 =
5.7 10^{-10}$ m$^2$/s. The simulations with HI result in
a scaling exponent of $m = -0.63 \pm 0.01$, which is in good agreement with
value obtained from experiments, $m = -0.64$ \cite{boehme03-07}, and a
previously reported result, $m = -0.617$, by \cite{stellwagen03a}. 
Only for the very short chains, ($N<5$), is a deviation from the prediction
observed and a higher diffusion coefficient found in the simulations.
For intermediate chain length, the coarse-grained simulation model with
HI, reproduces the experimentally observed behavior.

Without HI, the chains show the expected Rouse diffusion with an
exponent of $m = 1.02 \pm 0.02$. This simple model is clearly not applicable to
mimic the experimental behavior of short PE chains.

\begin{figure}[htp]
\begin{center}
  \includegraphics[width=\columnwidth]{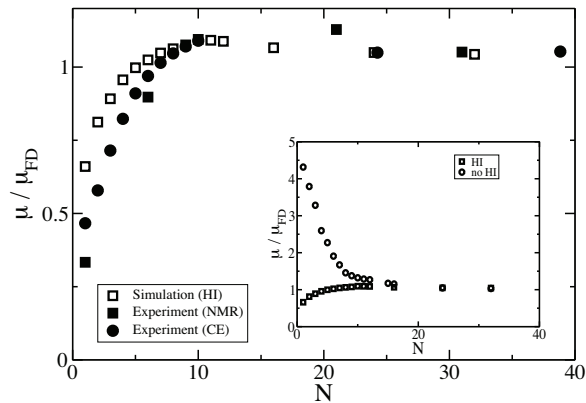}
  \caption[]{\label{fig:mobility}Normalized electrophoretic mobility
    $\mu / \mu_{\text{FD}}$ as a function of the number of repeat
    units $N$ for simulation with HI, from capillary electrophoresis
    (CE), and from electrophoretic NMR.  The inset compares
    simulations with and without HI.}
\end{center}
\end{figure}
 
\textbf{Electrophoretic mobility ($\mu$).} The results for
measurements of $\mu$ in pure water without additional salt are shown
in Fig. \ref{fig:mobility}. To account for the different viscosity of
the solvents, we rescale the mobilities by the free-draining mobility
$\mu_\text{FD}$ as obtained for long chains.

The experimental data sets agree
within the accuracies of the individual methods and show the characteristic behavior of the
mobility in dependence to chain length. A mobility maximum for $N=10$ is
observed with capillary electrophoresis. 
This maximum for intermediate
chains as well as the long chain
behavior is successfully reproduced in simulations with HI. For the first few oligomers, we
observe a small difference which is in line with the deviation for the diffusion.

On the other hand, as illustrated in the inset of Fig. \ref{fig:mobility}, the
simulation without hydrodynamic interactions fails completely to describe the
short chain
behavior and can only be mapped to the experimental data in the long chain
limit. Therefore, we infer that the mobility maximum can only be explained when
taking into account HI between the PE and
the surrounding solvent \footnote{Similar effects were found by R. Winkler using
SRD to model hydrodynamics; private communication.}.

\begin{figure}[htp]
\begin{center}
  \includegraphics[width=\columnwidth]{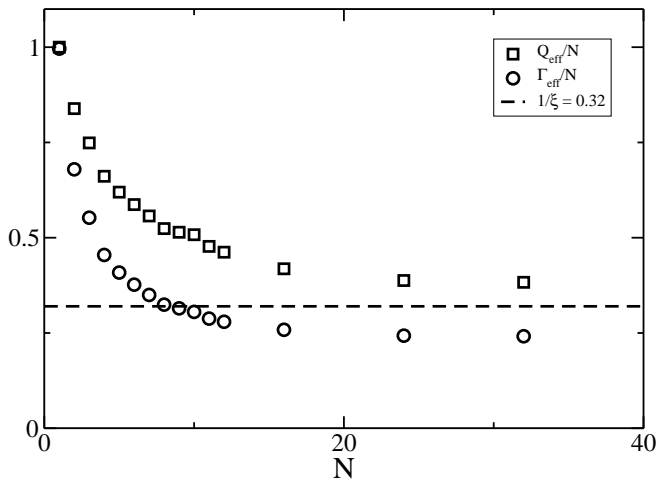}
  \caption{Effective charge per monomer $Q_\text{eff}/N$ (squares) as estimated
  from counterion condensation and effective friction per monomer
  $\Gamma_\text{eff}/N$. $Q_\text{eff}$ is approaching
  $Q/N=1/\xi=0.32$ as predicted by Manning theory.}
  \label{fig:xieff}
\end{center}
\end{figure}

To further illustrate this, we estimate the effective hydrodynamic
friction, $\Gamma_\text{eff}$, and the effective charge, $Q_\text{eff}$,
in dependence of its length. During electrophoresis, the electrical
driving force $F_E = Q_\text{eff} E$ is balanced by the frictional
force with the solvent $F_F = \Gamma_\text{eff} v$, where $v$ is the
migration velocity induced by the electric field $E$. For the
mobility, we then obtain: $\mu =
{Q_\text{eff}}/{\Gamma_\text{eff}}$. The PE is surrounded
by oppositely charged counter ions, some of which moving with the
chain and thus reducing its effective charge. We estimate this charge
reduction by subtracting the number of counter ions that are found
within 2 units of the chain from the bare charge $N$. This estimate is
used together with the obtained mobility to calculate the effective
friction as shown in Fig.  \ref{fig:xieff}. The effective charge for
long chains is in agreement with the Manning prediction at infinite
dilution, yielding $Q/N = 1/\xi = 0.32$. From Fig. \ref{fig:xieff}, we
see the impact of HI that results in the
mobility maximum.  Additional monomers are partly shielded from the
flow by the other monomers decreasing the effective friction per
monomer with chain length. This shielding is effective on short length
scales and leads to a stronger initial decrease of the friction than
the counter ion condensation reduces the effective charge. This leads
to an increasing mobility for intermediate chains. While the effective
friction levels off quickly, the effective charge per monomer keeps
decreasing slowly, reducing the mobility and causing the maximum. For
long chains, effective charge and effective friction per monomer
become constant, leading to the well-known free-draining behavior.

\section{Conclusion}

We performed a detailed study investigating the dynamic behavior of
short PE chains via MD simulations of a coarse grained
model and via two different experimental approaches. The results of
experiments and simulations can be quantitatively matched and agree
with the existing theory and predictions, as long as the simulation
model correctly includes long-range HI. A
simulation model that neglects HI fails to
reproduce the short-length scale behavior of the diffusion
coefficient and of the electrophoretic mobility.

To our knowledge we demonstrated for the first time, that the
transport coefficient of short PEs can be \emph{quantitatively}
modeled by coarse grained simulation techniques. No chemical details
are needed to explain the experimental results. Our model allows to
simulate time scales otherwise out of reach for atomistic simulations
and provides a microscopic understanding of the observed maximum in
$\mu$. From our observation we conclude, that hydrodynamic shielding
between the chain monomers is the major cause for this previously
little understood behavior, and that it is vitally important to
include HI to model such systems.
 
Having a simulation model at hand, that confirms the experimental
data, opens new possibilities of investigating the electrophoretic
behavior of short PEs, which so far has not been fully
explained by the existing theories.


\begin{acknowledgments}
  We thank B.~D\"{u}nweg, U.~Schiller, and G. Slater for helpful
  remarks.  Funds from the the Volkswagen foundation, the DAAD, and
  DFG under the TR6 are gratefully acknowledged.
\end{acknowledgments}


\begin{thebibliography}{22}
\expandafter\ifx\csname natexlab\endcsname\relax\def\natexlab#1{#1}\fi
\expandafter\ifx\csname bibnamefont\endcsname\relax
  \def\bibnamefont#1{#1}\fi
\expandafter\ifx\csname bibfnamefont\endcsname\relax
  \def\bibfnamefont#1{#1}\fi
\expandafter\ifx\csname citenamefont\endcsname\relax
  \def\citenamefont#1{#1}\fi
\expandafter\ifx\csname url\endcsname\relax
  \def\url#1{\texttt{#1}}\fi
\expandafter\ifx\csname urlprefix\endcsname\relax\def\urlprefix{URL }\fi
\providecommand{\bibinfo}[2]{#2}
\providecommand{\eprint}[2][]{\url{#2}}

\bibitem[{\citenamefont{Righetti}(1996)}]{righetti96a}
\bibinfo{editor}{\bibfnamefont{P.~G.} \bibnamefont{Righetti}}, ed.,
  \emph{\bibinfo{title}{Capillary Electrophoresis in Analytical Biotechnology}}
  (\bibinfo{publisher}{CRC Press, Boca Raton}, \bibinfo{year}{1996}).

\bibitem[{\citenamefont{Dolnik}(2006)}]{dolnik06a}
\bibinfo{author}{\bibfnamefont{V.}~\bibnamefont{Dolnik}},
  \bibinfo{journal}{Electrophoresis} \textbf{\bibinfo{volume}{27}},
  \bibinfo{pages}{126} (\bibinfo{year}{2006}).

\bibitem[{\citenamefont{Cottet et~al.}(2005)\citenamefont{Cottet, Simo,
  Vayaboury, and Cifuentes}}]{cottet05a}
\bibinfo{author}{\bibfnamefont{H.}~\bibnamefont{Cottet}},
  \bibinfo{author}{\bibfnamefont{C.}~\bibnamefont{Simo}},
  \bibinfo{author}{\bibfnamefont{W.}~\bibnamefont{Vayaboury}},
  \bibnamefont{and}
  \bibinfo{author}{\bibfnamefont{A.}~\bibnamefont{Cifuentes}},
  \bibinfo{journal}{Journal Of Chromatography A}
  \textbf{\bibinfo{volume}{1068}}, \bibinfo{pages}{59} (\bibinfo{year}{2005}).

\bibitem[{\citenamefont{Cottet and Gareil}(2007)}]{cottet07a}
\bibinfo{author}{\bibfnamefont{H.}~\bibnamefont{Cottet}} \bibnamefont{and}
  \bibinfo{author}{\bibfnamefont{P.}~\bibnamefont{Gareil}},
  \emph{\bibinfo{title}{CE from small ions to Macromolecules}}
  (\bibinfo{publisher}{Humana Press, NJ, USA}, \bibinfo{year}{2007}).

\bibitem[{\citenamefont{Nkodo et~al.}(2001)\citenamefont{Nkodo, Garnier,
  Tinland, Ren, Desruisseaux, McCormick, Drouin, and Slater}}]{nkodo01a}
\bibinfo{author}{\bibfnamefont{A.~E.} \bibnamefont{Nkodo et~al.}},
  \bibinfo{journal}{Electrophoresis} \textbf{\bibinfo{volume}{22}},
  \bibinfo{pages}{2424} (\bibinfo{year}{2001}).

\bibitem[{\citenamefont{Stellwagen and Stellwagen}(2002)}]{stellwagen02a}
\bibinfo{author}{\bibfnamefont{E.}~\bibnamefont{Stellwagen}} \bibnamefont{and}
  \bibinfo{author}{\bibfnamefont{N.~C.} \bibnamefont{Stellwagen}},
  \bibinfo{journal}{Electrophoresis} \textbf{\bibinfo{volume}{23}},
  \bibinfo{pages}{2794} (\bibinfo{year}{2002}).

\bibitem[{\citenamefont{Stejskal and Tanner}(1965)}]{stejskal65a}
\bibinfo{author}{\bibfnamefont{E.~O.} \bibnamefont{Stejskal}} \bibnamefont{and}
  \bibinfo{author}{\bibfnamefont{J.~E.} \bibnamefont{Tanner}},
  \bibinfo{journal}{Journal Of Chemical Physics} \textbf{\bibinfo{volume}{42}},
  \bibinfo{pages}{288} (\bibinfo{year}{1965}).

\bibitem[{\citenamefont{Stilbs and Furo}(2006)}]{stilbs06a}
\bibinfo{author}{\bibfnamefont{P.}~\bibnamefont{Stilbs}} \bibnamefont{and}
  \bibinfo{author}{\bibfnamefont{I.}~\bibnamefont{Furo}},
  \bibinfo{journal}{Current Opinion In Colloid \& Interface Science}
  \textbf{\bibinfo{volume}{11}}, \bibinfo{pages}{3} (\bibinfo{year}{2006}).

\bibitem[{\citenamefont{Scheler}(2002)}]{scheler02a}
\bibinfo{author}{\bibfnamefont{U.}~\bibnamefont{Scheler}},
  \emph{\bibinfo{title}{Handbook of Polyelectrolytes and their , Vol. 2}}
  (\bibinfo{publisher}{Tripaty, S. K. and Kumar, J. and Nalwa, H. S.},
  \bibinfo{year}{2002}), p. \bibinfo{pages}{173ff}.

\bibitem[{\citenamefont{B{\"o}hme and Scheler}(2003)}]{boehme03-07}
\bibinfo{author}{\bibfnamefont{U.}~\bibnamefont{B{\"o}hme}} \bibnamefont{and}
  \bibinfo{author}{\bibfnamefont{U.}~\bibnamefont{Scheler}},
  \bibinfo{journal}{Colloids and Surfaces A} \textbf{\bibinfo{volume}{222}},
  \bibinfo{pages}{35} (\bibinfo{year}{2003});
  \bibinfo{journal}{Journal Of Colloid And Interface Science}
  \textbf{\bibinfo{volume}{309}}, \bibinfo{pages}{231}
  (\bibinfo{year}{2007}{\natexlab{a}});
  \bibinfo{journal}{Macromolecular Chemistry and Physics}
  \textbf{\bibinfo{volume}{208}}, \bibinfo{pages}{2254} (\bibinfo{year}{2007}).

\bibitem[{\citenamefont{Hoagland et~al.}(1999)\citenamefont{Hoagland,
  Arvanitidou, and Welch}}]{hoagland99a}
\bibinfo{author}{\bibfnamefont{D.~A.} \bibnamefont{Hoagland}},
  \bibinfo{author}{\bibfnamefont{E.}~\bibnamefont{Arvanitidou}},
  \bibnamefont{and} \bibinfo{author}{\bibfnamefont{C.}~\bibnamefont{Welch}},
  \bibinfo{journal}{Macromolecules} \textbf{\bibinfo{volume}{32}},
  \bibinfo{pages}{6180} (\bibinfo{year}{1999}).

\bibitem[{\citenamefont{Cottet et~al.}(2000)\citenamefont{Cottet, Gareil,
  Theodoly, and Williams}}]{cottet00b}
\bibinfo{author}{\bibfnamefont{H.}~\bibnamefont{Cottet}},
  \bibinfo{author}{\bibfnamefont{P.}~\bibnamefont{Gareil}},
  \bibinfo{author}{\bibfnamefont{O.}~\bibnamefont{Theodoly}}, \bibnamefont{and}
  \bibinfo{author}{\bibfnamefont{C.~E.} \bibnamefont{Williams}},
  \bibinfo{journal}{Electrophoresis} \textbf{\bibinfo{volume}{21}},
  \bibinfo{pages}{3529} (\bibinfo{year}{2000}).

\bibitem[{\citenamefont{Stellwagen et~al.}(2003)\citenamefont{Stellwagen, Lu,
  and Stellwagen}}]{stellwagen03a}
\bibinfo{author}{\bibfnamefont{E.}~\bibnamefont{Stellwagen}},
  \bibinfo{author}{\bibfnamefont{Y.~J.} \bibnamefont{Lu}}, \bibnamefont{and}
  \bibinfo{author}{\bibfnamefont{N.~C.} \bibnamefont{Stellwagen}},
  \bibinfo{journal}{Biochemistry} \textbf{\bibinfo{volume}{42}},
  \bibinfo{pages}{11745} (\bibinfo{year}{2003}).

\bibitem[{\citenamefont{Muthukumar}(1996)}]{muthukumar96a}
\bibinfo{author}{\bibfnamefont{M.}~\bibnamefont{Muthukumar}},
  \bibinfo{journal}{Electrophoresis} \textbf{\bibinfo{volume}{17}},
  \bibinfo{pages}{1167} (\bibinfo{year}{1996}).

\bibitem[{\citenamefont{Volkel and Noolandi}(1995)}]{volkel95b}
\bibinfo{author}{\bibfnamefont{A.~R.} \bibnamefont{Volkel}} \bibnamefont{and}
  \bibinfo{author}{\bibfnamefont{J.}~\bibnamefont{Noolandi}},
  \bibinfo{journal}{Journal Of Chemical Physics}
  \textbf{\bibinfo{volume}{102}}, \bibinfo{pages}{5506} (\bibinfo{year}{1995}).

\bibitem[{\citenamefont{Mohanty and Stellwagen}(1999)}]{mohanty99a}
\bibinfo{author}{\bibfnamefont{U.}~\bibnamefont{Mohanty}} \bibnamefont{and}
  \bibinfo{author}{\bibfnamefont{N.~C.} \bibnamefont{Stellwagen}},
  \bibinfo{journal}{Biopolymers} \textbf{\bibinfo{volume}{49}},
  \bibinfo{pages}{209} (\bibinfo{year}{1999}).

\bibitem[{\citenamefont{Limbach et~al.}(2006)\citenamefont{Limbach, Arnold,
  Mann, and Holm}}]{limbach06a}
\bibinfo{author}{\bibfnamefont{H.~J.} \bibnamefont{Limbach}},
  \bibinfo{author}{\bibfnamefont{A.}~\bibnamefont{Arnold}},
  \bibinfo{author}{\bibfnamefont{B.~A.} \bibnamefont{Mann}}, \bibnamefont{and}
  \bibinfo{author}{\bibfnamefont{C.}~\bibnamefont{Holm}},
  \bibinfo{journal}{Comput. Phys. Commun.} \textbf{\bibinfo{volume}{174}},
  \bibinfo{pages}{704} (\bibinfo{year}{2006}).

\bibitem[{\citenamefont{Ahlrichs and D{\"u}nweg}(1999)}]{ahlrichs99a}
\bibinfo{author}{\bibfnamefont{P.}~\bibnamefont{Ahlrichs}} \bibnamefont{and}
  \bibinfo{author}{\bibfnamefont{B.}~\bibnamefont{D{\"u}nweg}},
  \bibinfo{journal}{J. Chem. Phys.} \textbf{\bibinfo{volume}{111}},
  \bibinfo{pages}{8225} (\bibinfo{year}{1999}).

\bibitem[{\citenamefont{Lobaskin et~al.}(2007)\citenamefont{Lobaskin, D\"unweg,
  Medebach, Palberg, and Holm}}]{lobaskin07a}
\bibinfo{author}{\bibfnamefont{V.}~\bibnamefont{Lobaskin}},
  \bibinfo{author}{\bibfnamefont{B.}~\bibnamefont{D\"unweg}},
  \bibinfo{author}{\bibfnamefont{M.}~\bibnamefont{Medebach}},
  \bibinfo{author}{\bibfnamefont{T.}~\bibnamefont{Palberg}}, \bibnamefont{and}
  \bibinfo{author}{\bibfnamefont{C.}~\bibnamefont{Holm}},
  \bibinfo{journal}{Phys. Rev. Lett.} \textbf{\bibinfo{volume}{98}},
  \bibinfo{pages}{176105} (\bibinfo{year}{2007}).

\bibitem[{\citenamefont{Gottwald et~al.}(2003)\citenamefont{Gottwald, Kuran,
  and Scheler}}]{gottwald03a}
\bibinfo{author}{\bibfnamefont{A.}~\bibnamefont{Gottwald}},
  \bibinfo{author}{\bibfnamefont{P.}~\bibnamefont{Kuran}}, \bibnamefont{and}
  \bibinfo{author}{\bibfnamefont{U.}~\bibnamefont{Scheler}},
  \bibinfo{journal}{Journal Of Magnetic Resonance}
  \textbf{\bibinfo{volume}{162}}, \bibinfo{pages}{364} (\bibinfo{year}{2003}).

\end{thebibliography}

\end{document}